\documentstyle{mn}
\input{psfig.sty}

\newif\ifAMStwofonts

\def\aj{{AJ}}			
\def\araa{{ARA\&A}}		
\def\apj{{ApJ}}			
\def\apjl{{ApJ}}		
\def\apjs{{ApJS}}

\def\aap{{A\&A}}

\def\mnras{{MNRAS}}

\def\fcp{{Fund.~Cosmic~Phys.}}

\ifoldfss
  \ifCUPmtlplainloaded \else
    \NewTextAlphabet{textbfit} {cmbxti10} {}
    \NewTextAlphabet{textbfss} {cmssbx10} {}
    \NewMathAlphabet{mathbfit} {cmbxti10} {} 
    \NewMathAlphabet{mathbfss} {cmssbx10} {} 
  \fi
  \ifAMStwofonts
    \ifCUPmtlplainloaded \else
      \NewSymbolFont{upmath} {eurm10}
      \NewSymbolFont{AMSa} {msam10}
      \NewMathSymbol{\upi}     {0}{upmath}{19}
      \NewMathSymbol{\umu}     {0}{upmath}{16}
      \NewMathSymbol{\upartial}{0}{upmath}{40}
      \NewMathSymbol{\leqslant}{3}{AMSa}{36}
      \NewMathSymbol{\geqslant}{3}{AMSa}{3E}

       \let\le=\leqslant
       
    \fi
  \fi
\fi 

\ifnfssone
  \newmathalphabet{\mathit}
  \addtoversion{normal}{\mathit}{cmr}{m}{it}
  \addtoversion{bold}{\mathit}{cmr}{bx}{it}
  \newmathalphabet{\mathbfit} 
  \addtoversion{normal}{\mathbfit}{cmr}{bx}{it}
  \addtoversion{bold}{\mathbfit}{cmr}{bx}{it}
  \newmathalphabet{\mathbfss} 
  \addtoversion{normal}{\mathbfss}{cmss}{bx}{n}
  \addtoversion{bold}{\mathbfss}{cmss}{bx}{n}
  \ifAMStwofonts
    \ifCUPmtlplainloaded \else
      \UseAMStwoboldmath
      \makeatletter
      \new@mathgroup\upmath@group
      \define@mathgroup\mv@normal\upmath@group{eur}{m}{n}
      \define@mathgroup\mv@bold\upmath@group{eur}{b}{n}
      \edef\UPM{\hexnumber\upmath@group}
      \new@mathgroup\amsa@group
      \define@mathgroup\mv@normal\amsa@group{msa}{m}{n}
      \define@mathgroup\mv@bold\amsa@group{msa}{m}{n}
      \edef\AMSa{\hexnumber\amsa@group}
      \makeatother
      \mathchardef\upi="0\UPM19
      \mathchardef\umu="0\UPM16
      \mathchardef\upartial="0\UPM40
      \mathchardef\leqslant="3\AMSa36
      \mathchardef\geqslant="3\AMSa3E

       \let\le=\leqslant

    \fi
  \fi
\fi 

\ifnfsstwo
  \DeclareMathAlphabet{\mathbfit}{OT1}{cmr}{bx}{it}
  \SetMathAlphabet\mathbfit{bold}{OT1}{cmr}{bx}{it}
  \DeclareMathAlphabet{\mathbfss}{OT1}{cmss}{bx}{n}
  \SetMathAlphabet\mathbfss{bold}{OT1}{cmss}{bx}{n}
  \ifAMStwofonts
    \ifCUPmtlplainloaded \else
      \DeclareSymbolFont{UPM}{U}{eur}{m}{n}
      \SetSymbolFont{UPM}{bold}{U}{eur}{b}{n}
      \DeclareSymbolFont{AMSa}{U}{msa}{m}{n}
      \DeclareMathSymbol{\upi}{0}{UPM}{"19}
      \DeclareMathSymbol{\umu}{0}{UPM}{"16}
      \DeclareMathSymbol{\upartial}{0}{UPM}{"40}
      \DeclareMathSymbol{\leqslant}{3}{AMSa}{"36}
      \DeclareMathSymbol{\geqslant}{3}{AMSa}{"3E}

       \let\le=\leqslant

    \fi
  \fi
\fi 

\ifCUPmtlplainloaded \else
  \ifAMStwofonts \else 
    \def\upi{\pi}
    \def\umu{\mu}
    \def\upartial{\partial}
  \fi
\fi

\title[The K-band Hubble diagram for the brightest cluster galaxies]
{The K-band Hubble diagram for the brightest cluster galaxies: 
a test of hierarchical galaxy formation models.}
\author[A. Arag\'on-Salamanca, C.M. Baugh and 
        G. Kauffmann]
       {Alfonso Arag\'on-Salamanca,$^1$ Carlton M. Baugh$^2$ and 
        Guinevere Kauffmann$^3$ \\
        $^1$Institute of Astronomy, Madingley Road, Cambridge CB3 0HA, England\\
        $^2$Department of Physics, Science Laboratories, South Road, 
            Durham DH1 3LE, England\\
        $^3$Max-Plank-Institut f\"ur Astrophysik, D-85740 Garching 
            bei M\"unchen, Germany 
        }

\date{Accepted ---.
      Received ---;
      in original form ---}

\pagerange{\pageref{firstpage}--\pageref{lastpage}}
\pubyear{---}

\begin{document}

\maketitle

\label{firstpage}

\begin{abstract}

We analyse the $K$-band Hubble diagram for a
sample of brightest cluster galaxies (BCGs) in the redshift range
$0<z<1$. In  good agreement with earlier studies, we confirm that the
scatter in the absolute magnitudes of the galaxies is  small
($0.3\,$magnitudes).  The BCGs exhibit very little luminosity evolution
in this redshift range:  if $q_0=0.0$ we detect {\it no\/} luminosity
evolution; for $q_0=0.5$ we measure a small {\it negative\/} evolution
(i.e., BCGs were about $0.5\,$magnitudes fainter at $z=1$ than today).
If the mass in stars of these galaxies had remained constant over this
period of time, substantial positive luminosity evolution would be
expected:  BCGs should have been {\it brighter\/} in the past since
their stars were younger. A likely explanation for the observed zero or
negative evolution is that the stellar mass of the BCGs has been
assembled over time through merging and accretion, as expected in
hierarchical models of galaxy formation.  The colour evolution of the
BCGs is consistent with that of an old stellar population ($z_{form} >2
$) that is evolving passively. We can thus use evolutionary population
synthesis models to estimate the rate of growth in stellar mass for
these systems. We find that the stellar mass in a typical BCG has grown
by a factor $\simeq 2$ since $z\simeq1$ if $q_0=0.0$ or by factor
$\simeq4$ if $q_0=0.5$. These results are in good agreement with the
predictions of semi-analytic models of galaxy formation and evolution
set in the context of a hierarchical scenario for structure formation.
The models predict a scatter in the luminosities of the BCGs
that is somewhat larger than the observed one, but that depends
on the criterion used to select the model clusters.

\end{abstract}

\begin{keywords}
galaxies: clusters --- galaxies: formation --- galaxies: evolution --- galaxies: elliptical and lenticular, cD --- infrared: galaxies
\end{keywords}

\section{Introduction}

Brightest cluster galaxies (BCGs) have been extensively studied at
optical wavelengths (\cite{Peach70}, 1972; \cite{GunnOke75},
\cite{SandageKW76}; \cite{KristianSW78}; \cite{KristianSW78};
\cite{Hoessel80}; \cite{SchneiderGH83a},b; \cite{LauerP92};
\cite{PostmanL95}). The small scatter in their absolute magnitudes
($<0.3\,$magnitudes up to  $z\sim 0.5$, see e.g. \cite{Sandage88}) and
their high luminosities make them useful standard candles for classical
cosmological tests involving the Hubble diagram, such as the
determination of $q_0$ and  the variation of $H_0$ with redshift.
However, there is now firm evidence that significant evolution has
taken place in the colours and optical luminosities of early-type
galaxies (including the BCGs) from $z\simeq0$ to $z\simeq1$ (e.g.
\cite{AragonECC93}; \cite{RakosS95}; \cite{OkeGH96}; \cite{Lubin96};
\cite{StanfordED97}) implying that the Hubble diagram could be
seriously affected by evolutionary changes even at relatively modest
redshifts. Conclusions concerning $q_0$ cannot be derived from such
diagrams until these changes are well-understood.  Indeed, the Hubble
diagram for the BCGs may well have more to teach us about galaxy
evolution than about cosmology.

The study of the BCGs in the near-infrared $K$-band ($2.2\mu$)  has two
main advantages over optical studies: first, the $k$-corrections are
appreciably smaller (indeed, they are negative) and virtually
independent of the spectral type of the galaxies (see, e.g.,
\cite{AragonECC93}, 1994); and second, the light at long wavelengths is
dominated by long-lived stars. Thus the $K$-band luminosity is a good
measure of the total stellar mass in the galaxies.

The $K$-band Hubble diagram has long been used as an evolutionary
diagnostic for radio galaxies (\cite{Grasdalen80}; \cite{Lebofsky80};
\cite{LebofskyE86}; \cite{LillyL82}, 1984; \cite{Lilly89a},b), which
also have a relatively small intrinsic dispersion in their
luminosities  ($\simeq0.4\,$ magnitudes for $1<z<2.2$;
\cite{Lilly89a}).  Substantial positive luminosity evolution was
found:  radio galaxies were about one magnitude brighter at $z\simeq1$
than they are today.  Arag\'on-Salamanca et al.  (1993) measured
$K$-band luminosities for a sample of~19~BCGs in the redshift range 
$0<z<1$  and found that the $K$ magnitude-redshift relation is tighter
(scatter$\,=0.3\,$magnitudes) than for radio galaxies.  Moreover, these
authors did not detect significant $K$ luminosity evolution for the
BCGs, and concluded that their evolutionary properties were different
from those of radio galaxies:  powerful radio galaxies are apparently
less homogeneous in their evolutionary behaviour and show  stronger
luminosity evolution.  Arag\'on-Salamanca et al. also found that the
colours of the early-type galaxies (including the BCGs) in these
clusters have evolved  since $z\simeq1$ at a rate consistent with a
scenario in which  the bulk of their stars formed at relatively high
redshifts ($z>2$) and evolved passively thereafter.  This conclusion
has been confirmed by several more recent studies (\cite{RakosS95};
\cite{OkeGH96}; \cite{Lubin96}; \cite{EllisEA97};
\cite{StanfordED97}).

The recent development of {\it semi-analytic techniques\/} has provided
theorists with the tools to make predictions for the formation and
evolution of galaxies, using physically motivated models set in the
context of hierarchical structure formation in the universe (e.g.,
\cite{WhiteF91}; \cite{Cole91}; \cite{LaceyS91}; \cite{KauffmannWG93};
\cite{KauffmannGW94}; \cite{ColeAFNZ94}; \cite{Heyl95}).  The
properties of galaxies in these models are in broad agreement with the
present day characteristics of galaxies, such as the distribution of
luminosities, colours and morphologies and with the properties of
galaxies at high redshift, including the faint galaxy counts, colours
and redshift distributions (e.g. \cite{Kauffmann95};
\cite{Kauffmann96}; \cite{KC97}; \cite{BaughCF96a};
\cite{BaughCFL97}).  In this paper, we will test whether the models
also predict the right behaviour for a special class of galaxies,
namely the BCGs. In the models, these galaxies grow in mass both from
the cooling of gas from the surrounding hot halo medium and from the
accretion of ``satellite'' galaxies that fall to the cluster centre as
result of dynamical friction and then merge.

In section~2 we present the observational results and derive the
luminosity evolution of the BCGs in the sample. In section~3 we briefly
describe two independent semi-analytic models of galaxy formation,
outline their basic assumptions, limitations and uncertainties, and
discuss the predictions they make for the evolution of the BCGs.
Section~4 compares the model predictions with the observational
results.

\section{Observational results}

\subsection{Photometric data}

The clusters in our sample are all optically-selected, although most of
them have also been detected in X-rays.  At $z\le0.37$ they come from
the Abell (1957) and Abell, Corwin \& Olowin (1989) catalogues.  At
higher redshifts they have been found from the projected density
contrast in deep optical imaging surveys (\cite{GunnHO86};
\cite{CouchEMM91}) and should represent the richest clusters present at
each redshift.

The $K$-band data analysed here come from the photometric study carried
out by Arag\'on-Salamanca et al. (1993) of galaxy clusters in the
$0<z<1$ redshift range (19 BCGs) complemented with new $K$-band images
of~3~clusters at $z\simeq3$ (\cite{BargerEA96}) and~6 clusters with
$0.39\le z\le 0.56$ (\cite{BargerEA97}).  The images were obtained at
the 3.8-m UK Infrared Telescope, the 3.9-m Anglo-Australian Telescope
and the Palomar Observatory 1.52-m Telescope.  There is some overlap in
the clusters observed by Arag\'on-Salamanca et al.  and Barger et al.,
bringing the number of BCGs studied here to a total of 25 (see
Table~\ref{tbl1}).  For the clusters in common, both sets of photometry
agree very well (within the estimated errors) and we have chosen the
better quality image in each case.

\begin{table*}
\centering
\begin{minipage}{8.5cm}
\begin{center}
\caption{Corrected rest-frame $K$-band magnitudes for the BCGs \label{tbl1} } 
\begin{tabular}{lcccc}
\hline
\hline
{Cluster}      & 
{$z$} &
{$K^0_{{\rm q}_0=0.0}$} &
{$K^0_{{\rm q}_0=0.5}$} &
{reference} \\
\hline
Abell 1656$^{\rm a}$(NGC4889) & 0.023 & \phantom{1}8.89 & \phantom{1}8.88 & (1) \\
Abell 2199 (NGC6166) & 0.030 & \phantom{1}9.50   &  \phantom{1}9.49 & (1) \\
Abell 2197 (NGC6173) & 0.031 & \phantom{1}9.54  &  \phantom{1}9.53 & (1) \\
Abell 2151 (NGC6034) & 0.037 & 10.49 & 10.38 & (1) \\
Abell 963            & 0.206 & 13.61 & 13.55 & (1) \\
Abell 1942           & 0.224 & 13.92 & 13.88 & (1) \\
AC103                & 0.310 & 14.62 & 14.56 & (2) \\
AC114                & 0.310 & 14.54 & 14.46 & (2) \\
AC118                & 0.310 & 14.77 & 14.67 & (2) \\
Cl2244$-$02          & 0.329 & 15.35 & 15.31 & (1) \\
Abell 370            & 0.374 & 15.05 & 14.95 & (1) \\
Cl0024$+$16          & 0.391 & 15.29 & 15.14 & (1),(3) \\
3C295                & 0.460 & 15.26 & 15.14 & (3) \\
Cl0412$-$65$^{\rm b}$ & 0.510 & 15.60 & 15.55 & (3) \\
Cl1601$+$42          & 0.540 & 16.23 & 16.13 & (3) \\
Cl0016$+$16          & 0.546 & 16.21 & 16.02 & (1),(3) \\
Cl0054$-$27$^{\rm c}$  & 0.563 & 16.50 & 16.37 & (1),(3) \\
Cl0317$+$1521        & 0.583 & 17.08 & 17.01 & (1) \\
Cl0844$+$18$^{\rm d}$ & 0.664 & 16.93 & 16.80 & (1) \\
Cl1322$+$3029        & 0.697 & 16.90 & 16.72 & (1) \\
Cl0020$+$0407        & 0.698 & 17.06 & 17.01 & (1) \\
Cl1322$+$3027        & 0.751 & 16.97 & 16.79 & (1) \\
Cl2155$+$0334        & 0.820 & 17.22 & 17.12 & (1) \\
Cl1603$+$4313        & 0.895 & 17.78 & 17.64 & (1) \\
Cl1603$+$4329        & 0.920 & 18.24 & 18.02 & (1) \\
\hline
\end{tabular}
\end{center}
$^{\rm a}${Coma cluster.}\\
$^{\rm b}${Also known as F1557.19TC (Couch et al. 1991).}\\
$^{\rm c}${Also known as J1888.16CL (Couch et al. 1991).}\\
$^{\rm d}${Also known as F1767.10TC (Couch et al. 1991).}\\

References. --- {(1) Arag\'on-Salamanca et al. 1993;   (2) Barger et
al. 1996; (3) Barger et al. 1997}

\end{minipage}
\end{table*}

We refer the reader to the original papers for a detailed description
of the data. We will concentrate here on the aspects relevant to this
paper. The photometric zero-point uncertainties in the $K$-band
images were always small ($0.02$--$0.04\,$magnitudes) and are
completely negligible for our analysis.  As in Arag\'on-Salamanca et
al. (1993), $K$ photometry for the BCGs was obtained inside a fixed
metric aperture of 50$\,$kpc diameter\footnote{$H_0 =
50\,$km$\,$s$^{-1}\,$Mpc$^{-1}$ assumed throughout.  Because of the
uncertainty in the value of $q_0$ we will carry out parallel analyses
for $q_0=0.0$ and $0.5$} so that the minimum aperture at high redshift
is $\sim5\,$arcsec thus minimising the effect of seeing.  Because of
the ambiguity of what constitutes a ``galaxy'' ($\sim30\%$ of local
first-ranked cluster galaxies are multiple-nucleus systems:  i.e., more
than one ``galaxy'' occurs within  our metric aperture ---see
\cite{Hoessel80}) we follow the approach of Schneider, Gunn \& Hoessel
(1983a) and adopt a working definition of the brightest cluster galaxy
as {\it the region of maximum cluster light enclosed in our metric
aperture}. The photometry data from the AAT and PO was transformed into
the $K$-band system of the UKIRT (where most of the data comes from)
using colour terms determined from the known filter responses. The
colour terms were always $\le0.08\,$magnitudes, and we estimate that
the uncertainty in the transformation is well below $2\%$ and thus
negligible.  Galactic reddening corrections were estimated from
Burstein \& Heiles' (1982) maps, but since the galactic latitudes were
always reasonably high, they were very small in the $K$-band. Finally,
small seeing corrections and $k$-corrections were also applied as in
Arag\'on-Salamanca et al.  Table~\ref{tbl1} contains the corrected
$K$-magnitudes for the BCG sample.  Values for $q_0=0.0$ and $0.5$ are
given to account for the difference in projected aperture at a given
metric aperture as a function of the deceleration parameter. We
estimate that the total uncertainty of the corrected magnitudes is
below 10\%,  and will contribute very little to the observed scatter
(see original data papers for a detailed discussion of the
uncertainties).

\begin{figure}
\psfig{figure=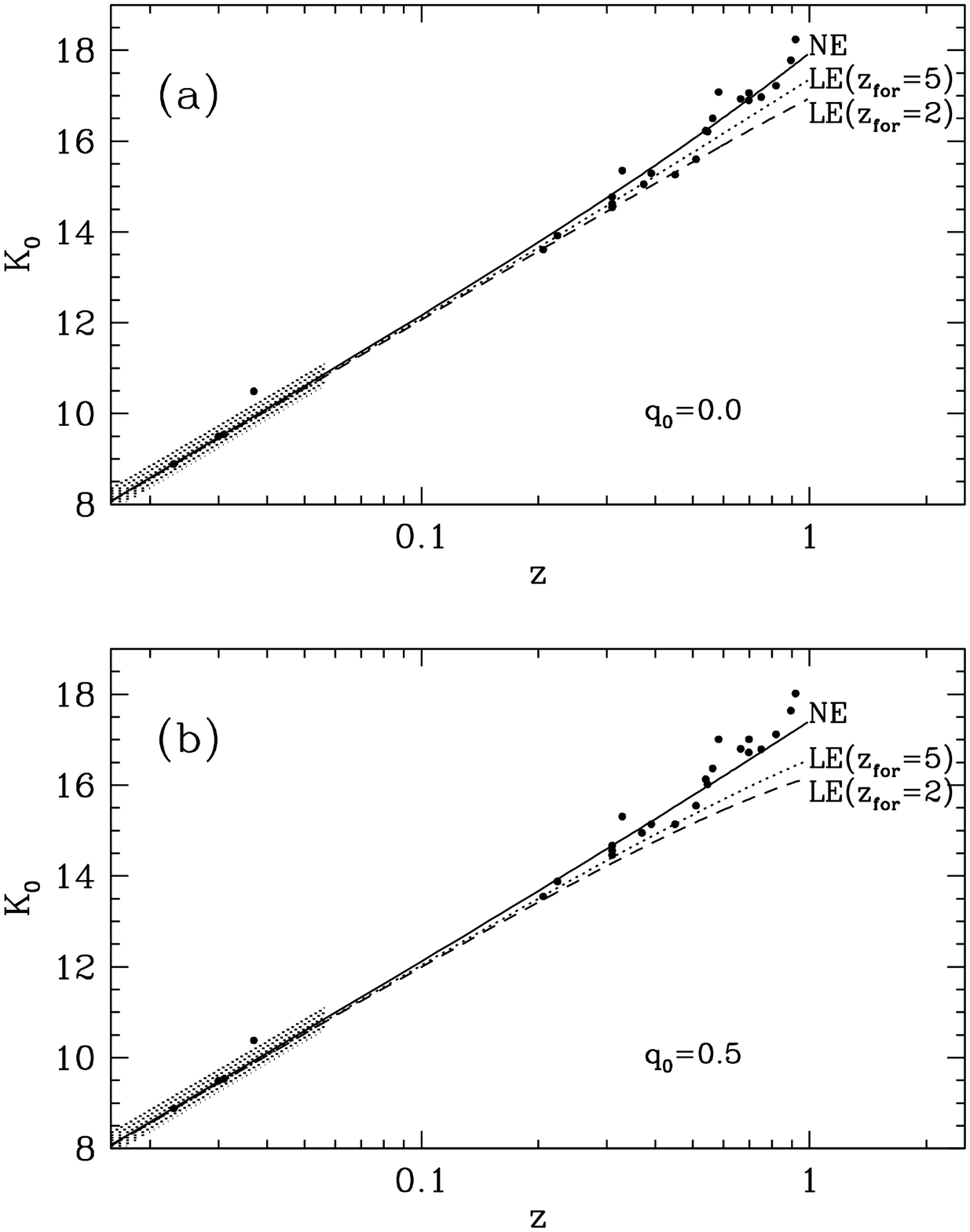,width=85mm,height=140mm}
\caption{{\bf (a)} Magnitude-redshift relation
(Hubble diagram) for the Brightest Cluster Galaxies in the rest-frame
$K$-band.  The $K$ magnitudes have been measured inside a projected
50$\,$kpc-diameter aperture.  The solid line shows the no-evolution
prediction, normalised to the low-redshift data. The dashed line
corresponds to the luminosity evolution expected for a stellar
population formed at $z_{\rm for} = 2$ evolving passively.  The dotted
line shows the evolution expected for $z_{\rm for} = 5$.  The shaded
area at low-redshift represents the  $K$-band magnitudes of $z<0.06$
BCGs estimated from the $R$-band data of Postman \& Lauer. The width of
the shaded area reflects the observed scatter (see text for details).
$H_0 = 50\,$km$\,$s$^{-1}\,$Mpc$^{-1}$ and $q_0=0.0$ were assumed.
{\bf (b)} As (a) but for $q_0=0.5$.  \label{fig1} }
\end{figure}

Figure~\ref{fig1} shows the $K$ magnitude--redshift relation (Hubble
diagram) for the BCGs in our sample. The solid lines show no-evolution
predictions which only take into account the distance modulus as a
function of $z$, normalised to the three lowest redshift points. Since
we do not have $K$ magnitudes for many low redshift galaxies, the
normalisation of the no-evolution line could be somewhat uncertain.
However, extensive optical photometry exists and, given the homogeneity
in the colours of these galaxies (\cite{PostmanL95}), we feel justified
to use the $R$-band photometry of Lauer \& Postman (1992) to estimate
$K$-band magnitudes.  The shaded area at low-redshift in
Figure~\ref{fig1} represents the $K$-band magnitudes for $z<0.06$ BCGs
estimated from the $R$-band data of Postman \& Lauer, corrected to our
metric aperture (using $d\log L/d\log r = 0.5$; from the same authors),
and assuming a typical $R-K=2.6$ (cf. \cite{BowerLE92a},b). The width
of the shaded area reflects the observed scatter. The agreement with
our local normalisation is remarkably good.

In agreement with the results of Arag\'on-Salamanca et al., the
scatter of the observed magnitudes around the no-evolution prediction
is $0.30\,$mag.  Assuming that the $K$-band light provides an estimate
of the total stellar mass of the galaxies, this very low scatter
implies that BCGs at a given redshift have very similar masses in stars
(within 30\% {\it r.m.s.}), which has yet to be explained by any model
of galaxy formation (see section~3).

Morphological information is available for the low redshift clusters
from the ground-based images. Many of intermediate and high redshift
clusters have been imaged with HST, thus morphological information
exists for 16 of the $z>0.3$ clusters (\cite{SmailEA96}, 1997;
\cite{CouchEA97}). In general, the BCGs are either cD galaxies or giant
ellipticals, although for the highest-$z$ clusters there could be some
ambiguity since the extended cD halos might not be clearly visible
given the relatively high surface brightness limits achievable with
HST. Thus some of the BCGs classified as ellipticals could be cD
galaxies. The exception to this rule is the BCG in the 3C295 cluster,
which is a radio galaxy and shows an un-resolved AGN-like morphology in
a disturbed disk or envelope.

The optical and optical-infrared colours (from ground-based and HST
data) of the BCGs are compatible with those of the colour-magnitude
sequence of cluster ellipticals and S0s. The exception is, again,
3C295, which is substantially bluer (by $\simeq0.8\,$mag in $R-K$).
The peculiar colours and morphology of this galaxy are probably related
to its being a powerful radio source.  However, its $K$-band luminosity
agrees very well with that of the rest of the BCGs in our sample, and
we have kept it in our analysis.  Taking it out would not alter our
conclusions.

\subsection{Interpretation using evolutionary synthesis models}

\begin{figure}
\psfig{figure=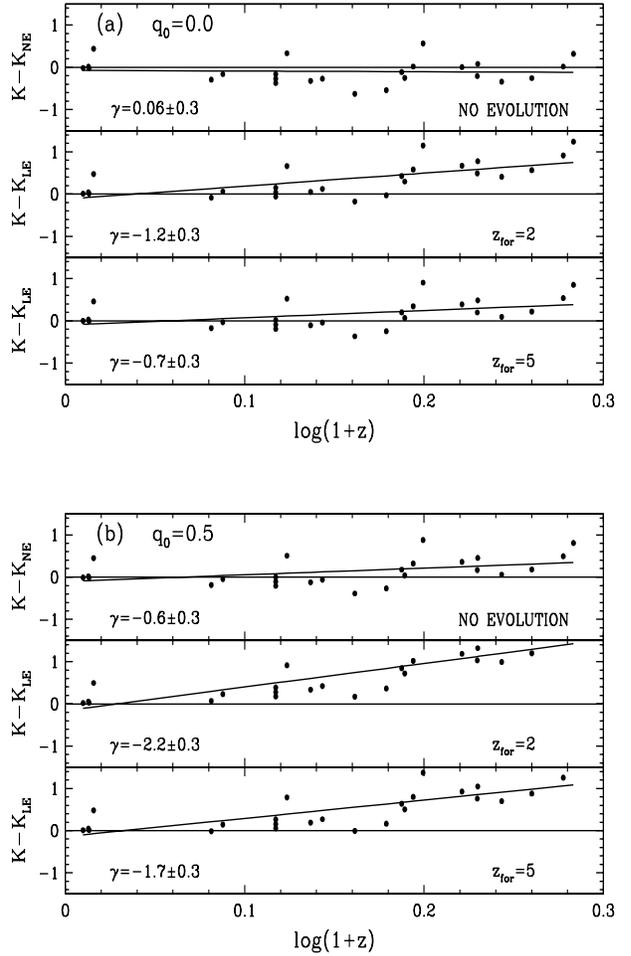,width=85mm,height=140mm}
\caption{{\bf (a)} Top panel: The same data presented in
figure~\ref{fig1} after subtracting the no-evolution prediction. Middle
and bottom panels:  the same data after subtracting models for
luminosity evolution in which the BCG stars form at $z_{\rm for} = 2$
and $z_{\rm for} = 5$ respectively and evolve passively thereafter (for
$q_0=0.0$). The solid lines represent least-squares linear fits.  {\bf
(b)} As (a) but for $q_0=0.5$. \label{fig2} }
\end{figure}

Figure~\ref{fig1} shows that the luminosity of the BCGs (inside a fixed
metric aperture of $50\,$kpc diameter) does not evolve strongly with
redshift. This is shown more clearly on the top panel of
Figure~\ref{fig2}, where the no-evolution prediction has been
subtracted from the data. It is clear that for $q_0=0.0$ the observed
magnitudes do not seem to evolve with redshift.  For $q_0=0.5$ there is
a hint of {\it negative\/} evolution:  BCGs at high redshift tend to be
marginally fainter than low redshift ones.

A convenient parametrisation of the luminosity evolution is
$L_K(z)=L_K(0)\times(1+z)^\gamma$. Least-squares fits to the data yield
$\gamma =-0.06\pm0.3$ and $-0.6\pm0.3$, for $q_0 = 0.0$ and $0.5$
respectively. Thus the BCGs show {\it no\/} or marginally {\it
negative\/} luminosity evolution. However, as mentioned above, the
colours of these galaxies show the same evolution as the early-type
cluster galaxies: they get progressively bluer with redshift at a rate
which indicates that their stellar populations were formed at $z>2$ and
evolved passively thereafter. If the total stellar mass of the galaxies
has remained constant, we would expect them to get progressively {\it
brighter\/} with redshift, as the average ages of their stellar
populations get younger. Since that brightening is not observed, the
most likely explanation is that the total mass in stars in the BCGs has
not remained constant but has grown with time. We will now estimate the
rate of this growth.

We used evolutionary population synthesis models (\cite{BruzualC97};
see also \cite{BruzualC93}) to predict the expected luminosity
evolution of a passively-evolving stellar population formed at a given
redshift.  That should represent the brightening of the stellar
populations due to the decrease in average stellar age with redshift. 
A Scalo (1986) Initial Mass Function with stellar masses in
the range $0.1 M_\odot\le m\le 125M_\odot$ was assumed. The exact value
of the upper mass limit is not critical given the expected ages of the
stellar populations in the observed redshift range, provided that it is
above the turn-off mass at the ages corresponding to $z<1$.  The model
predictions are plotted on Figure~\ref{fig1} (LE lines) for formation
redshifts $z_{\rm for} = 2$ and  $z_{\rm for} = 5$.  These models
produce a colour evolution compatible with the observations of
early-type cluster galaxies. Note that $z_{\rm for}<2$ is clearly ruled
out by the observed {\it slow\/} colour evolution (cf. section~1).
Making $z_{\rm for}>5$ makes very little difference to the age of the
stars.  We have assumed solar metallicity, but changing the metal
content does not significantly alter the predictions. Other models
(e.g. \cite{ArimotoY86,KodamaA97}) predict very
similar luminosity evolution,  thus we believe that the models
considered here bracket reasonably well the luminosity evolution
expected from the colour constraints and that the predictions are not
very model-dependent.

In Figure~\ref{fig2} (lower two panels) we have plotted the observed
magnitudes after subtracting the luminosity evolution predictions.
Since the models assume a constant stellar mass, the rate of change
with redshift in the $K$ magnitudes, after correcting for the expected
brightening due to passive evolution, should be a direct measurement of
the rate of change in stellar mass.  Note that strictly speaking, what
we call ``mass in stars'' really means ``mass in luminous stars'',
i.e., those which contribute to the galaxy $K$-band light.
Parameterising the evolution as
$M_{\star}(z)=M_{\star}(0)\times(1+z)^\gamma$, least-squares fits give
the following results:
$$
\matrix{
q_0 = 0.0 & z_{\rm for}=2 & \gamma=-1.2\pm0.3 \cr
            & z_{\rm for}=5 & \gamma=-0.7\pm0.3 \cr
q_0 = 0.5 & z_{\rm for}=2 & \gamma=-2.2\pm0.3 \cr
            & z_{\rm for}=5 & \gamma=-1.7\pm0.3 \cr
}
$$

These rates imply that the mass in stars of a typical BCG grew by 
a factor $\simeq2$ if $q_0 = 0.0$ or $\simeq4$ if $q_0 = 0.5$ from
$z\simeq1$ to $z\simeq0$, i.e., in the last $\simeq8$--$10\,$Gyrs.

At this point we would like to make a historical comment. One of us
(AAS) carried out the analysis described in this section {\it
independently\/} of the rest of the authors and invited the Durham and
Munich groups to make predictions, using the semi-analytic models described 
in section~3, for the growth of the stellar mass in BCGs as a function of
redshift. In the first instance, this was done  {\it without any 
prior knowledge of the observationally determined rates} and without any 
interaction between the two groups to ensure that the predictions where 
truly independent. The surprising level of agreement of the model 
predictions with each other, and with the observations,    
encouraged us to compare the 
model output with the data in more detail. This is presented in the next 
section. 

\section{Galaxy formation and evolution models}

In this section, we discuss the predictions of the semi-analytic models
of the Durham and Munich groups. Full details of the Durham models can
be found in \cite{ColeAFNZ94} and \cite{BaughCF96b}.  The version of
the Munich models used here is the same as that described in
\cite{KC97}.  Further details about the Munich models can be found in
\cite{KauffmannWG93}.  Although the two models are very similar in
outline and in their basic framework, many features, for example the
simple parameterisations used to describe star formation and feedback,
are different in detail.

Briefly, the hierarchical collapse and merging of an ensemble of dark
matter halos is followed using Monte Carlo techniques.  Gas virialises
in the dark matter halos,  cools radiatively and condenses onto a {\it
central\/} galaxy at the core of each halo.  Star formation occurs at a
rate proportional to the mass of cold gas present. The supply of cold
gas is regulated by feedback processes associated with star formation,
such as stellar winds and supernovae.  As time proceeds, a halo will
merge with a number of others, forming a new halo of larger mass. All
gas which has not already cooled is assumed to be shock heated to the
virial temperature of this new halo. This hot gas then cools onto the
central galaxy of the new halo, which is identified with the central
galaxy of its {\it largest progenitor}. The central galaxies of the
other progenitors become {\it satellite galaxies}, which are able to
merge with the central galaxy on a dynamical friction timescale. If  a
merger takes place between two galaxies of roughly comparable mass, the
merger remnant is labelled as an ``elliptical'' and all cold gas is
transformed instantaneously into stars in a ``starburst''.  The same
stellar population models (\cite{BruzualC97}) used in Section~2.2 are
employed to turn the star formation histories of the model galaxies
into broad band luminosities.

In these models, there are three different ways in which the stellar
masses of the brightest galaxies in clusters can grow:
\begin {enumerate}
\item merging of satellite galaxies that sink to the center of the halo
through dynamical friction
\item quiescent star formation as a result of gas cooling 
from the surrounding hot halo medium
\item ``bursts'' of star formation associated with the accretion of a massive satellite.
\end {enumerate}

As noted by \cite{KauffmannWG93}, the colours and absolute
magnitudes of central cluster galaxies are predicted to be bluer and
brighter than observed if all the gas present in the cooling flows of massive
clusters turns into visible stars. To fix this problem, these authors
simply switched off star formation in cooling flows when the circular
velocity of the halo exceeded $500\,$km$\,$s$^{-1}$. 
This value was chosen so as to obtain a good fit to the 
bright end of the Virgo cluster luminosity function.
Another solution is to assume that  gas in the central regions of halos
follows a different density profile to that of the dark matter.
In particular, if the gas has a constant density core, cooling is very much
less efficient in the centres of massive clusters.
The precise form of the gas density profile is much less important for cooling
in low mass halos, since these objects are denser and have much shorter cooling
times.

\begin{figure}
\psfig{figure=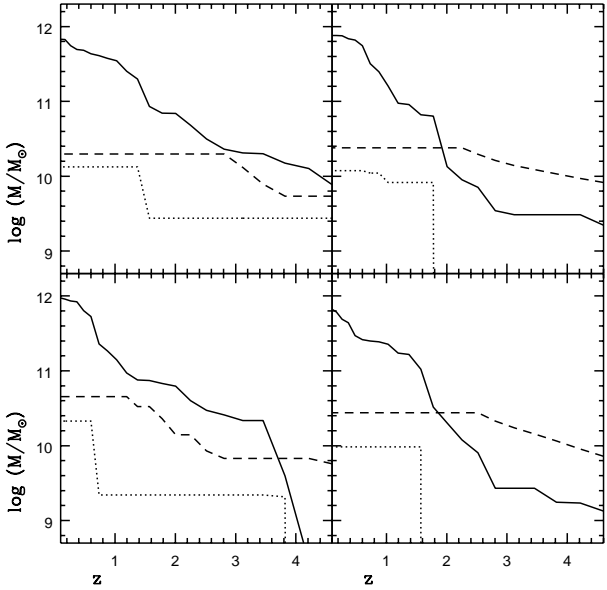,width=85mm}
\caption{
The build up in stellar mass of the largest progenitor in four 
examples of BCGs taken from the Munich models, which reside in dark 
matter halos with circular velocity $1000 {\rm kms}^{-1}$ at the 
present day. Star formation is switched off in any progenitor halo whose 
circular velocity is in excess of $500 {\rm kms}^{-1}$.
The solid lines show the accumulation of stellar mass from merging 
events, the dashed lines show the mass contributed by stars forming 
from gas cooling from the surrounding hot
halo medium, and the dotted lines show the mass contributed by star
formation bursts associated with major mergers.  
\label{fighist}}
\end{figure}

In figure~\ref{fighist}, we consider the evolution of a subset of the
$z=0$ BCGs in a model with star formation switched off in halos with 
circular velocity greater than $500\,$km$\,$s$^{-1}$. The solid line shows
the cumulative growth in stellar mass of the BCG due to merging
events.  The dashed line shows the cumulative contribution from stars
formed from quiescently cooling gas, and the dotted line is the
contribution from stars formed in bursts during major mergers. As can
be seen, star formation from bursts and cooling gas account for only a
few percent of the final mass of the BCG. The rest comes from accreted
galaxies. The four panels in the figure represent independent Monte
Carlo realizations of the formation of a BCG in a halo of circular
velocity $1000\,$km$\,$s$^{-1}$.  It is striking that although each BCG
has a significantly different merging history, their final stellar
masses differ very little from one realization to another.

\begin{figure}
\psfig{figure=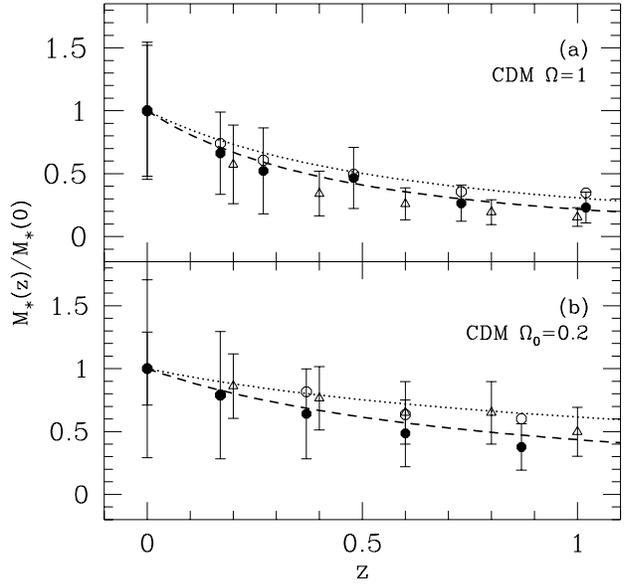,width=90mm}
\caption{
The average stellar mass of BCGs as a function of redshift.  We have
selected galaxies at each epoch that reside in halos with circular
velocities greater than $700\,$km$\,$s$^{-1}$.  The points have been
normalised by dividing by the average stellar mass of the BCGs in each
model at $z=0$.  (a) shows the predictions for a Cold Dark Matter
universe with the critical density, whilst (b) shows an open universe
with a present day density parameter of $\Omega_{0}=0.2$.  In each
panel, the filled circles show the results from the Munich models and
the open traingles show the results of the Durham models.  The curves
show the trends in stellar mass deduced in Section 2.2; the dotted
curve corresponds to $z_{for}=5$ and the dashed curve to $z_{for}=2$.
The open circles in (a) and (b) show the evolution in the average
stellar mass of the BCGs predicted by the Munich model when a cut in
halo mass of $2 \times 10^{14} h^{-1} M_{\odot}$ is applied to select
clusters at each redshift.
\label{figmass}}
\end{figure}

\begin{figure}
\psfig{figure=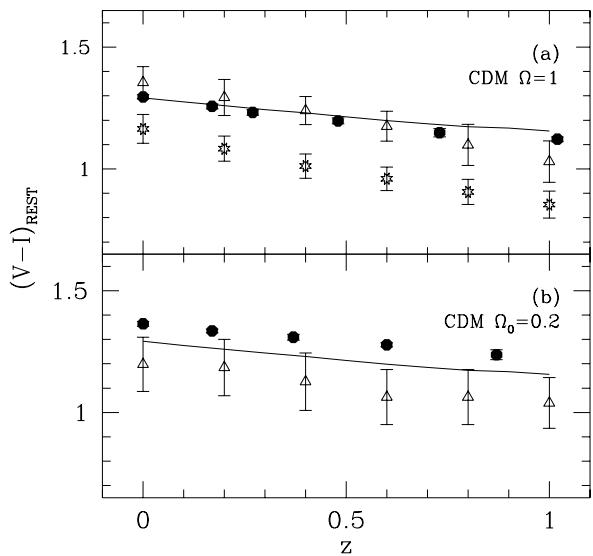,width=85mm}
\caption{
The rest frame $V-I$ colour of the model BCGs as a function 
of redshift. Again, (a) shows the results for a critical density 
CDM universe, whilst (b) shows the model output for an open universe 
with $\Omega_{0}=0.2$.
The filled symbols show the Munich models and the open symbols 
show the Durham models. 
The open stars in (a) show a Durham model in which the gas 
density profile is the same as that of the dark matter in the halo.
Hence, gas cools more efficiently and more recent star formation takes place,  
making the BCG's bluer. 
The line is a fit to the observed colour evolution of the BCG's taken from  
Arag\'on-Salamanca et al. (1993). 
\label{figvir}}
\end{figure}

\begin{figure}
\psfig{figure=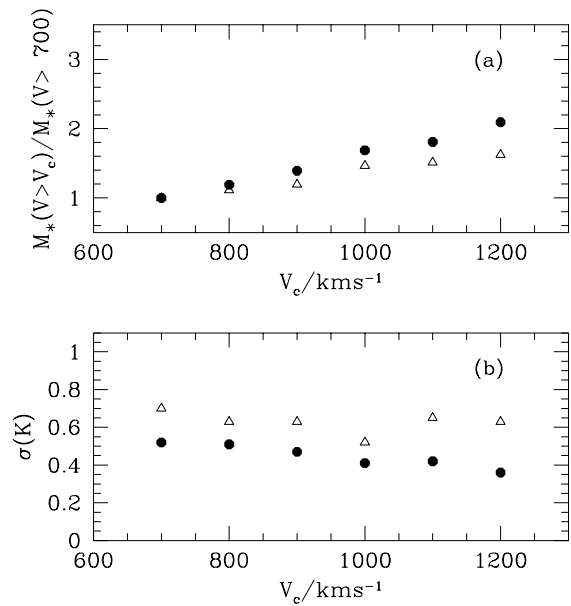,width=90mm}
\caption{
(a) The average stellar mass of BCGs at $z=0$ as a function of 
the circular velocity cut used to select clusters. 
The mean BCG mass increases by a factor of two over a range in 
circular velocity that corresponds to a increase by a factor of 
5 in dark matter halo mass. The filled symbols show the Munich results 
and the open symbols show the Durham results.
(b) The scatter in the absolute K-magnitudes of BCGs as a function of 
the circular velocity used to define the cluster sample. The Durham 
models predict a larger scatter than the Munich models.
\label{ksigma}}
\end{figure}

We will now make a comparison with the data presented in section 2.
The observational selection picks out {\it the richest clusters\/} at
each redshift, but the sample is not complete in any well-defined sense.
We mimic the observational selection as best we can  by calculating the masses of
BCGs in halos with circular velocities greater than some fixed value as a function of
redshift. This means we are selecting rarer objects with increasing
redshift. We  also explore the effect of selecting clusters by mass instead
of circular velocity.

We present the model predictions for two Cold Dark Matter cosmologies;
one with the critical density and an open model with density parameter
$\Omega_{0} = 0.2$. The density fluctuations in each case are
normalised to reproduce the abundance of rich clusters (e.g.
\cite{whiteEF93}; \cite{ekeCF96}; \cite{vianaL96}).  At several
redshifts in the range $0 \le z \le 1$, we have selected $\sim 200$ halos with
circular velocities greater than $700\,$km$\,$s$^{-1}$. Note that the 
clusters are weighted so that they are representative of the mass distribution 
of halos at the given redshift.

The evolution of the average stellar mass of the brightest galaxies in the halos with
redshift is shown in Figure~\ref{figmass}.  We have divided the
average mass of the BCGs at each redshift by the average mass of BCGs
found in the redshift zero clusters.  The open triangles show the results of
the Durham models, which adopt the halo density profile of \cite{NFW} for the
dark matter.
The gas is assumed to follow a different density profile, with a constant 
density at the core of the halo (full details of these extensions to the 
original \cite{ColeAFNZ94} model will be given in Cole et al, in preparation).  
The filled circles show the predictions of the Munich models, 
in which isothermal density profiles are assumed for both the gas 
and the dark matter. 
Star formation is switched off in halos with a circular velocity
in excess of $500\,$km$\,$s$^{-1}$.  
The curves show the growth in stellar mass with redshift of the BCGs
deduced in Section~2.2; the dotted lines show the parameterisation with
$z_{for}=5$ whilst the dashed line shows $z_{for}=2$. Both models fit
the observed evolution very well.

In Figure~\ref{figvir}, we show the evolution in the
average rest-frame $V-I$ colours of the BCGs. Once again, the open triangles
show the Durham models and the filled circles the Munich models. The curves
are a fit to the observed colour evolution given in \cite{AragonECC93}. 
For the $\Omega=1$ cosmology, both models fit the observed colour evolution
reasonably well. The Durham models predict larger scatter in the colours
because the BCGs are still undergoing some low-level star formation, whereas the
BCGs in the Munich models have ceased forming stars altogether. 
For reference, we also show a model 
(shown by the open stars) where  cooling and star formation are efficient 
at the centres of massive clusters.
This is a model where both the gas and the dark matter follow the standard \cite{NFW}
density profile.
In this case, the colours are too blue by $\sim 0.1-0.2$ magnitudes.  
In the low-density cosmology, it is interesting that the Durham and Munich models exhibit
{\it opposite} trends in colour. The Munich models are redder than in a flat cosmology,
because the BCGs form earlier and contain older stars. On the other hand, the Durham   
models are bluer. This is because halos in a low-density Universe are more concentrated    
and gas  still cools at the cluster centres, even if a different density profile
is assumed.

An important question to consider is how our results would be affected
by adopting a different selection criterion for the clusters in the
model. To explore this, Figure \ref{ksigma}a shows how the average mass
of a $z=0$ BCG changes as a function of circular velocity cutoff for
both the Munich and Durham models.  This plot shows that more massive
BCGs are found in more massive halos.  The stellar mass of a BCG
changes by a factor of 2 for a factor 5 increase in halo mass.  Figure
\ref{ksigma}b shows how the predicted scatter in K-band absolute
magnitude changes as a function of circular velocity cutoff.  As can be
seen, the scatter decreases as richer clusters are selected.  Both the
Munich and Durham models predict larger scatter than that observed,
though the discrepancy is worse for the Durham models.

It should be noted that clusters of fixed circular velocity are less
massive at high redshift by a factor $ \simeq (1+z)^{3/2}$ (this
scaling arises from the fact that halos collapse and virialize at a
density $\sim 200 \bar{\rho}(z)$, where $\bar {\rho}(z)$ is the mean
density of the Universe at redshift $z$).  Selecting clusters according
to a halo mass cutoff, rather than a circular velocity cutoff, will
thus select more massive BCGs at high redshift. The effect of mass
selection on the predicted evolution is shown by the open circles in
Figure~\ref{figmass}, where we have selected clusters greater than $2
\times 10^{14} M_{\odot}$ at each redshift.  The mass evolution is now
slightly weaker, but the effect is small and the model results are still in good
agreement with the observations.

\section{Discussion}

We have shown that the $K$-band Hubble diagram for a sample of
brightest cluster galaxies in the redshift range $0<z<1$ has a very
small scatter ($0.3\,$magnitudes), in  good agreement with earlier
studies. The BCGs exhibit very little luminosity evolution in this
redshift range:  if $q_0=0.0$ we detect {\it no\/} luminosity
evolution; for $q_0=0.5$ we measure a small {\it negative\/} evolution
(i.e., BCGs were about $0.5\,$magnitudes fainter at $z=1$ than today).
But substantial {\it positive\/} luminosity evolution would be expected
if the mass in stars of these galaxies had remained constant over this
period of time:  BCGs should have been {\it brighter\/} in the past
since their stars were younger.  The most likely explanation for the
observed zero or negative evolution is that the stellar mass of the
BCGs has been assembled over time through merging and accretion, as
expected in hierarchical models of galaxy formation.  Since the colour
evolution of the BCGs is consistent with that of an old stellar
population ($z_{form} >2$) that is evolving passively, we have used
evolutionary population synthesis models to estimate the rate of growth
in stellar mass for these systems. We find that the stellar mass in a
typical BCG has grown by a factor $\simeq 2$ since $z\simeq1$ if
$q_0=0.0$ or by factor $\simeq4$ if $q_0=0.5$.

Two independent groups of galaxy formation modellers predicted the
decrease in stellar mass of the BCGs with increasing redshift,
in advance of seeing the data and without any fine tuning of their
semi-analytic model parameters.  We have extracted the properties of
BCGs in massive halos with circular velocities greater than 700 km s$^{-1}$ or masses
greater than $2 \times 10^{14} M_{\odot}$, at various
redshifts.
The average stellar mass of the BCGs increases by a factor of $\sim 4-5$
in the critical density models.  In the low density models, the growth
in stellar mass is more modest, between a factor~$2$ to~$3$. This is
because clusters assemble at higher redshift in a low density Universe.
Note, however, that both models agree well with the data when the
appropriate value of $q_0$ is used in the analysis. It is not
possible to make statements about a preferred cosmology from this
exercise.

We also  find that in order to reproduce the observed colours of the
galaxies, some mechanism must act to suppress visible star formation in
the cooling flows of the most massive halos. It has been postulated
that much of this gas may be forming low-mass stars (\cite{fabian}) or
cold gas clouds (\cite{white}). Another possibility we have explored is
that the gas may be less concentrated in the centres of clusters than
the dark matter, and hence not able to cool efficiently. In both cases,
the evolution in the $K$-band luminosities of the BCGs comes about
almost exclusively  as a result of the merging of satellite galaxies in
the cluster.  An important and unanticipated result of our analysis is
that the predicted scatter in the stellar masses of the BCGs built
through merging and accretion is still quite small, although somewhat
larger than the observed one. The scatter depends on the criteria used
to select the model clusters.

\section*{Acknowledgments}

AAS acknowledges generous financial support from the Royal Society.  We
thank T.~Shanks for encouraging us to have a closer look at the
$K$-band Hubble diagram, and G.~Bruzual and S.~Charlot for allowing us
to use their model results prior to publication.  We thank the
anonymous referee for useful comments.  The version of the Durham
semi-analytic models used to produce the predictions in this paper is
the result of a collaboration between Shaun Cole, Carlos Frenk, Cedric
Lacey and CMB, supported in part by a Particle Physics and Astronomy
Research Council rolling grant. This work was carried out under the
auspices of EARA, a European Association for Research in Astronomy, and
the TMR Network on Galaxy Formation and Evolution funded by the
European Commission.

\bsp

\label{lastpage}


\begin{thebibliography}{}

   \bibitem[Abell 1957]{Abell57}
Abell, G.O. 1957, \apjs, 3, 221
   \bibitem[Abell, Corwin \& Olowin (1989)]{AbellCO89}
Abell, G.O., Corwin, H.G. \& Olowin, R.P. 1989, \apjs, 70, 1
   \bibitem[Arag\'on-Salamanca et al.\ 1993]{AragonECC93}
Arag\'on-Salamanca, A., Ellis, R.S., Couch, W.J. \& Carter, D. 1993, 
\mnras, 262, 764
   \bibitem[Arag\'on-Salamanca et al.\ 1994]{AragonESB94}
Arag\'on-Salamanca, A., Ellis, R.S., Schwartzenberg, J.-M. \& 
Bergeron, J. 1994, \mnras, 421, 27
   \bibitem[Arimoto \& Yoshii 1986]{ArimotoY86}
Arimoto, N. \& Yoshii, Y. 1986, \aap, 164, 260
   \bibitem[Barger et al. 1996]{BargerEA96} 
Barger, A.J., Arag\'on-Salamanca, A., Ellis, R.S., Couch, W.J., 
Smail, I. \& Sharples, R.M. 1996, \mnras, 279, 1
   \bibitem[Barger et al. 1997]{BargerEA97} 
Barger, A.J, Arag\'on-Salamanca, A., Smail, I., Ellis, R.S.,
Couch, W.J., Dressler, A., Oemler, A., Poggianti, B.M. \& Sharples, R.M. 1997, 
\apj, in press
   \bibitem[Baugh, Cole \& Frenk 1996a]{BaughCF96a}
Baugh, C.M., Cole, S. \& Frenk, C.S., 1996, \mnras, 283, L15
   \bibitem[Baugh, Cole \& Frenk 1996b]{BaughCF96b}
Baugh, C.M., Cole, S. \& Frenk, C.S., 1996b, \mnras, 283, 1361
   \bibitem[Baugh et. al. 1997]{BaughCFL97}
Baugh, C.M., Cole, S., Frenk, C.S. \& Lacey, C.G., 1997, \apj\  
submitted, astro-ph/9703111
   \bibitem[Bower, Lucey \& Ellis 1992a]{BowerLE92a}
Bower, R.G., Lucey, J.R. \& Ellis, R.S., 1992a, \mnras, 254, 589
   \bibitem[Bower, Lucey \& Ellis 1992b]{BowerLE92b}
Bower, R.G., Lucey, J.R. \& Ellis, R.S., 1992b, \mnras, 254, 601
   \bibitem[Bruzual \& Charlot 1993]{BruzualC93}
Bruzual A., G. \& Charlot, S. 1993, \apj, 405, 538
   \bibitem[Bruzual \& Charlot 1997]{BruzualC97}
Bruzual A., G. \& Charlot, S. 1997, in preparation
   \bibitem[Burstein \& Heiles 1982]{BursteinH82}
Burstein, D. \& Heiles, C. 1982, \aj, 87, 1165
   \bibitem[Cole 1991]{Cole91}
Cole, S., 1991, \apj, 367, 45
   \bibitem[Cole et al. 1994]{ColeAFNZ94}
Cole, S., Arag\'on-Salamanca, A., Frenk, C.S., Navarro, J.F. \& Zepf, S.E. 
1994, \mnras, 271, 781
   \bibitem[Couch et al. 1991]{CouchEMM91}
Couch, W.J., Ellis, R.S., Malin, D.F. \& MacLaren, I. 1991, \mnras, 249, 606
\bibitem[Couch et al. 1997]{CouchEA97}
Couch, W.J. et al. 1997, in preparation
\bibitem[Eke, Cole \& Frenk, 1996]{ekeCF96}
Eke, V.R., Cole, S., Frenk, C.S., 1996, \mnras, 282, 263
   \bibitem[Ellis  et al.\ 1997]{EllisEA97}
Ellis, R.S., Smail, I., Dressler, A., Oemler Jr., A., Butcher, H. \&
Sharples, R.M. 1997, \apj, 483, 582
\bibitem[Fabian, Nulsen \& Canizares, 1982]{fabian}
Fabian, A.C., Nulsen , P.E.J. \& Canizares, C.R., 1982, MNRAS, 201, 933   
\bibitem[Grasdalen 1980]{Grasdalen80}
Grasdalen, G.L. 1980, in Objects of High Redshift, IAU Symp. No. 92, ed. G.O. Abell \&  P.J.E. Peebles, Reidel, Dordrecht: Reidel, 269
   \bibitem[Gunn \& Oke 1975]{GunnOke75} 
Gunn, J.E. \& Oke, J.B. 1975, \apj, 195, 255
   \bibitem[Gunn, Hoessel \& Oke 1986]{GunnHO86}
Gunn, J.E., Hoessel, J.G. \& Oke, J.B. 1986, \apj, 306, 30
   \bibitem[Heyl et. al. 1995]{Heyl95}
Heyl, J.S., Cole, S., frenk, C.S., Navarro, J.F., 1995, \mnras, 
   274, 755
   \bibitem[Hoessel 1980]{Hoessel80} 
Hoessel, J.G. 1980, \apj, 241, 493
   \bibitem[Kauffmann 1995]{Kauffmann95}
Kauffmann, G., 1995, \mnras, 274, 161
   \bibitem[Kauffmann 1996]{Kauffmann96}
Kauffmann, G., 1996, \mnras, 281, 487
   \bibitem[Kauffmann, White \& Guiderdoni 1993]{KauffmannWG93}
Kauffmann, G., White, S.D.M. \& Guiderdoni, B. 1993, \mnras, 264, 201
   \bibitem[Kauffmann, Guiderdoni \& White 1994]{KauffmannGW94}
Kauffmann, G., Guiderdoni, B. \& White, S.D.M. 1994, \mnras, 267, 981
   \bibitem[Kauffmann \& Charlot 1997]{KC97}
Kauffmann, G., \& Charlot, S., 1997, \mnras,  
in press, astro-ph/9704148
   \bibitem[Kristian, Sandage \& Westphal 1978]{KristianSW78} 
Kristian, J. Sandage, A. \& Westphal, J.A. 1978, \apj, 221 383
   \bibitem[Kodama \& Arimoto 1997]{KodamaA97}
Kodama, T. \& Arimoto, N. 1997, \aap, 320, 41
   \bibitem[Lacey \& Silk 1991]{LaceyS91}
Lacey, C.G. \& Silk, J. 1991, \apj, 381, 14
   \bibitem[Lauer \& Postman 1992]{LauerP92} 
Lauer, T.R. \& Postman, M. 1992, \apjl, 400, L47
   \bibitem[Lebofsky 1980]{Lebofsky80}
Lebofsky, M.J. 1980, in Objects of High Redshift, IAU Symp. No. 92, ed. G.O. Abell \&  P.J.E. Peebles, Reidel, Dordrecht: Reidel, 257
   \bibitem[Lebofsky \& Eisenhardt 1986]{LebofskyE86} 
Lebofsky, M.J. \& Eisenhardt, P.R.M. 1986, \apj, 300, 151
   \bibitem[Lilly \& Longair 1982]{LillyL82} 
Lilly, S.J. \& Longair, M.S. 1982, \mnras, 199, 1053
   \bibitem[Lilly \& Longair 1984]{LillyL84} 
Lilly, S.J. \& Longair, M.S. 1984, \mnras, 211, 833
   \bibitem[Lilly 1989a]{Lilly89a}
Lilly, S.J., 1989a, \apj, 340, 77
   \bibitem[Lilly 1989a]{Lilly89b}
Lilly, S.J., 1989b, in The Epoch of Galaxy Formation, ed. C.S. Frenk et al.,\  Dordrecht: Kluwer, 63
   \bibitem[Lubin 1996]{Lubin96} 
Lubin, L.M. 1996, \aj, 112, 23
\bibitem[Navarro, Frenk \& White 1996]{NFW}
Navarro, J.F., Frenk, C.S. \& White, S.D.M., 1996, \apj, 462, 563
   \bibitem[Oke, Gunn \& Hoessel 1996]{OkeGH96}
Oke, J.B., Gunn, J.E. \& Hoessel, J.G. 1996, \aj, 111, 29
   \bibitem[Peach 1970]{Peach70} 
Peach, J.V. 1970, \apj 159, 753
   \bibitem[Peach 1972]{Peach72} 
Peach, J.V. 1972, in External Galaxies and Quasi-Stellar Objects, 
IAU Symp. No. 44, ed. D.S. Evans, Dordrecht: Reidel, 314
   \bibitem[Postman \& Lauer 1995]{PostmanL95}
Postman, M. \& Lauer, T.R. 1995, \apj, 440, 28
   \bibitem[Rakos \& Schombert 1995]{RakosS95} 
Rakos, K.D. \& Schombert, J.M. 1995, \apj, 439, 47
   \bibitem[Sandage 1988]{Sandage88} 
Sandage, A. 1988, \araa, 26, 561
   \bibitem[Sandage, Kristian \& Westphal 1976]{SandageKW76}
Sandage, A., Kristian, J. \& Westphal, J.A. 1976, \apj, 205,  688
   \bibitem[Scalo 1986]{Scalo86}
Scalo, J.M. 1986, \fcp, 11, 1
   \bibitem[Schneider, Gunn \& Hoessel 1983a]{SchneiderGH83a} 
Schneider, D.P., Gunn, J.E. \& Hoessel, J.G. 1983a, \apj, 264, 337
   \bibitem[Schneider, Gunn \& Hoessel 1983b]{SchneiderGH83b}  
Schneider, D.P., Gunn, J.E. \& Hoessel, J.G., 1983b, \apj, 268, 476
   \bibitem[Smail et al. 1996]{SmailEA96}
Smail, I, Dressler, A., Kneib, J.-P., Ellis, R.S., Couch, W.J., Sharples, R.M.
\& Oemler Jr., A. 1996, \apj, 469, 508
   \bibitem[Smail et al. 1997]{SmailEA97}
Smail, I, Dressler, A., Couch, W.J., Ellis, R.S., Oemler Jr., A., Butcher, H. 
\& Sharples, R.M. 1997, \apjs, 110, 213
   \bibitem[Stanford, Eisenhardt \& Dickinson 1997]{StanfordED97}
Stanford, S.A.,  Eisenhardt, P.R.M. \& Dickinson, M.E. 1997, in preparation
\bibitem[Viana \& Liddle 1996]{vianaL96}
Viana, P.T.P. \& Liddle, A.R., 1996, \mnras 281,
\bibitem[White et al. 1991]{white}
White, D.A., Fabian, A.C., Johnstone, R.M., Mushotsky, R.F. \&
Arnaud, K.A., 1991, MNRAS, 252, 72   
\bibitem[White \& Frenk 1991]{WhiteF91}
White, S.D.M., \& Frenk, C.S., 1991, \apj, 379, 52
\bibitem[White, Efstathiou \& Frenk 1993]{whiteEF93}
White, S.D.M., Efstathiou, G., \& Frenk, C.S., 1993, \mnras, 262, 1023

\end{thebibliography}
\end{document}